\documentclass{article}

\usepackage{arxiv}

\usepackage[authoryear]{natbib}
\usepackage{doi}
\usepackage{graphicx}
\usepackage{amsmath}
\usepackage{caption}

\usepackage{siunitx}
\usepackage{enumitem}
\usepackage{placeins} %floatbarrier
\usepackage{tikz}
\usetikzlibrary{er,arrows, chains, fit, positioning, calc, shapes}

\usepackage{booktabs}
\usepackage{longtable}
\usepackage{array}
\usepackage{multirow}
\usepackage{wrapfig}
\usepackage{float}
\usepackage{colortbl}
\usepackage{pdflscape}
\usepackage{tabu}
\usepackage{threeparttable}
\usepackage{threeparttablex}
\usepackage[normalem]{ulem}
\usepackage{makecell}
\usepackage{xcolor}
\usepackage{nowidow}
\usepackage{csquotes}
\MakeOuterQuote{"}

\title{Improving Adaptive Seamless Designs through Bayesian optimization}
\renewcommand{\shorttitle}{Bayesian optimization of Adaptive Seamless Designs}

\hypersetup{
pdftitle={Improving Adaptive Seamless Designs through Bayesian optimization},
pdfauthor={Jakob Richter, Tim Friede, Jörg Rahnenführer},
pdfkeywords={Adaptive Seamless Designs, Bayesian Optimization, Clinical Trials, Treatment Selection},
}

\author{
  \href{https://orcid.org/0000-0003-4481-5554}{\includegraphics[scale=0.06]{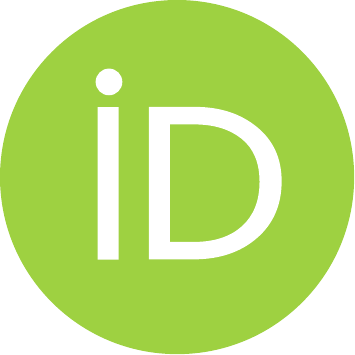}\hspace{1mm}Jakob Richter}  \\
    Fakultät Statistik \\
    Technische Universität Dortmund \\
    \texttt{richter@statistik.tu-dortmund.de}
  \And
  \href{https://orcid.org/0000-0001-5347-7441}{\includegraphics[scale=0.06]{orcid.pdf}\hspace{1mm}Tim Friede} \\
    Institut für Medizinische Statistik \\
    Universitätsmedizin Göttingen, and Deutsches Zentrum für Herz-Kreislauf-Forschung (DZHK), Standort Göttingen \\
    \texttt{tim.friede@med.uni-goettingen.de}
  \And
  \href{https://orcid.org/0000-0002-8947-440X}{\includegraphics[scale=0.06]{orcid.pdf}\hspace{1mm}Jörg Rahnenführer} \\
    Fakultät Statistik \\
    Technische Universität Dortmund \\
    \texttt{rahnenfuehrer@statistik.tu-dortmund.de}
}

\begin{document}
\maketitle

% max 250 words
\begin{abstract}
We propose to use Bayesian optimization (BO) to improve the efficiency of the design selection process in clinical trials.
BO is a method to optimize expensive black-box functions, by using a regression as a surrogate to guide the search.
In clinical trials, planning test procedures and sample sizes is a crucial task.
A common goal is to maximize the test power, given a set of treatments, corresponding effect sizes, and a total number of samples.
From a wide range of possible designs we aim to select the best one in a short time to allow quick decisions.
The standard approach to simulate the power for each single design can become too time-consuming.
When the number of possible designs becomes very large, either large computational resources are required or an exhaustive exploration of all possible designs takes too long.
Here, we propose to use BO to quickly find a clinical trial design with high power from a large number of candidate designs.
We demonstrate the effectiveness of our approach by optimizing the power of adaptive seamless designs for different sets of treatment effect sizes.
Comparing BO with an exhaustive evaluation of all candidate designs shows that BO finds competitive designs in a fraction of the time.
\end{abstract}

%% maketitle must follow the abstract.
\maketitle % Produces the title.

%% If there is not enough space inside the running head
%% for all authors including the title you may provide
%% the leftmark in one of the following three forms:

%% \renewcommand{\leftmark}
%% {First Author: A Short Title}

%% \renewcommand{\leftmark}
%% {First Author and Second Author: A Short Title}

%% \renewcommand{\leftmark}
%% {First Author et al.: A Short Title}

%% \tableofcontents  % Produces the table of contents.

\section{Introduction}

Clinical research and in particular drug development is typically structured into phases, e.g. phases I to IV in drug development. 
Similar approaches exist for complex interventions, such as the MRC framework for non-pharmacological interventions~\citep{campbell_framework_2000}.
The development phases include elements of learning and confirmation.
For instance, \citet{sheiner_learning_1997} describe the drug development process as two learning and confirming cycles:
The first cycle includes learning about the tolerated dose (Phase I) and confirming the efficacy of the selected dose in a selected group of patients (Phase IIa), whereas the second cycle consists of learning about the optimal use in representative patients (Phase IIb) and confirming an acceptable benefit or risk ratio (Phase III).

Traditionally, separate studies are performed for learning and confirming.
However, the landscape in clinical research is changing.
Boundaries between the development phases are increasingly dissolving in multiple aspects.
Master protocols are a new concept where multiple subgroups and substudies are integrated~\citep{bogin_master_2020}.
Here, simultaneously, more than one treatment and more than one disease type are investigated within the same overall trial structure.
Adaptive designs are established, but are being further developed.
In general, study designs become more flexible, and more principled approaches to decision-making using statistical modeling and integrated analyses across studies become popular. 
A~related principle on the analysis side is evidence synthesis, which is a way of combining information from several studies that have examined the same question to come to an overall understanding.

The increasing complexity of both the designs and the analyses of clinical trials has the consequence that the planning of such studies often relies also on computer simulations, especially if no closed form solutions for calculations are available.
This applies even to usually fairly straightforward tasks as sample size calculations. 
Various frameworks and guidance proposals have been developed over the past years on how to plan, execute and interpret the results of such simulation studies.
Method comparison studies are used to give clinicians and biostatisticians guidance when an approach improves currently used methods~\citep{hanneman_design_2008}.  
Depending on the setting and the purpose of the simulation study, suitable metrics must be chosen to compare alternative designs or analysis strategies. 
Especially relevant are neutral comparison studies, which are characterized by~\citet{boulesteix_plea_2013} as follows:
The primary goal of the respective article is not to introduce a new promising method, the authors are reasonably neutral, and the evaluation metrics, methods, and data sets are chosen in a rational way.
%While method comparison studies aim to make statements when one approach might dominate others in terms of the chosen metrics, in designing an individual study or series of studies the objective is to identify a design that is optimal in terms of the metrics chosen in a certain scenario or, more importantly, across a set of scenarios. 
A comprehensive framework for such an objective assessment of competing strategies for clinical trial designs is CSE (Clinical Scenario Evaluation)~\citep{benda_aspects_2010,friede_refinement_2010}.
CSE was particularly developed to support the overall process of performing simulations of clinical trials, especially for complex designs and analysis strategies.
The problem is decomposed into data models, analysis models, and evaluation models that specify the evaluation metrics.

Here, we propose to use a formal approach to optimization of the clinical trial design, in the situation of a very large number of available options for the parameters specifying the design. 
This set of candidate designs can be regarded as a multi-dimensional search space, including potentially many real-valued and categorical parameters.
The vast space of possible designs makes the optimization computationally demanding.
In the following, we demonstrate that the so-called \emph{Bayesian optimization} (BO) framework~\citep{jones_taxonomy_2001} can be successfully applied to select a clinical trial design.
As evaluation metric we consider the specific case that the statistical power of the clinical trial design should be maximized, for given effect sizes and a fixed total sample size.
Further, it is crucial that such a design is found in a short time.
%\subsection{Related work}
%subsub Clinical 
% TODO: Tim

BO, often also called Model-based optimization, has been successfully applied to optimize expensive black-boxes (i.e.\ functions that take long to evaluate and are not given in closed-form) in many scenarios.
Popular applications include hyper-parameter tuning of machine learning methods~\citep{snoek_practical_2012}, general algorithm configuration~\citep{hutter_sequential_2011}, and many more.
Various implementations of BO methods are widely used, such as SMAC~\citep{hutter_sequential_2011}, Spearmint~\citep{snoek_practical_2012}, and BoTorch~\citep{balandat_botorch_2020}.
For this work we use the implementation in the R-package \texttt{mlrMBO}~\citep{bischl_mlrmbo_2017}.
This implementation has been successfully used to optimize hyper-parameters of machine learning methods on various tasks~\citep{bischl_mlrmbo_2017, wozniak_candle_2018}, and in the biomedical context it has been applied to optimize model weights~\citep{richter_modelbased_2019,browaeys_nichenet_2020}.
Also, BO is well suited for multi-objective optimization.
Different adaptions of BO exist~\citep{horn_modelbased_2015} that return a set of non-dominated points.
Many of them are implemented in \texttt{mlrMBO} as well.
However, in this work we restrict ourselves to single-objective optimization. 
Independently of this work, for instance, \citet{wilson_efficient_2021} successfully applied multi-objective BO to minimize the number of participants and the number of clusters in a cluster randomized controlled trial under the restriction that the power is above a given threshold.

The remainder of the manuscript is structured as follows.
In the next section we present as motivating example an application of adaptive seamless designs for chronic obstructive pulmonary disease (COPD).
In the sections~\ref{sec:adaptive_seamless_designs} and~\ref{sec:bayesian_optimization}, we explain the background on adaptive seamless designs (asd) and Bayesian optimization (BO), respectively.
In Section~\ref{sec:simulation_study}, we present a simulation study that is closely related to the motivating example and in which the suitability of BO for optimizing trial designs is demonstrated.
In the last Section~\ref{sec:discussion}, we discuss promising extensions as well as limitations of the approach.

\section{Motivating Example}
\label{sec:motivating_example}

The general idea of this paper is to apply BO for finding a trial design that is optimal with respect to a certain metric.
As an example, we consider the maximization of the statistical power of a trial design, given a set of parameters to be chosen for the trial design.
We compare the computation time as well as the maximal obtained power between the results obtained with BO and with an exhaustive grid search over the space of adjustable parameters.
The presented optimization approach can also be applied for optimizing other metrics than the power, as long as the metric itself is numeric and the parameters of the trial design are numeric and include none or just few categorical choices.

One specific trial design that falls into this category are \emph{adaptive seamless designs}~\citep{barnes_integrating_2010} as implemented in the \texttt{asd} package~\citep{parsons_package_2012} and described in Section~\ref{sec:adaptive_seamless_designs}.
%(What is THE asd paper? peter bauer, meinter kieser, 1999 statistics in medicine, % https://doi.org/10.1002/(SICI)1097-0258(19990730)18:14%3C1833::AID-SIM221%3E3.0.CO;2-3, )  bauer_combining_1999
% https://doi.org/10.1002/bimj.200510232 bretz_confirmatory_2006 
% seamless trials (assoziert) 
%  https://doi.org/10.1002/bimj.200510231 schmidli_confirmatory_2006
In~\citet{friede_adaptive_2020} the authors present an application of \emph{adaptive seamless designs} for chronic obstructive pulmonary disease (COPD).
We use this COPD study to apply BO for finding a trial design with maximal statistical power, given the parameters of the respective design options.
%within the given scenarios restrictions.

%phase II III klinische studien je in Phase
% nicht 100% II = stage 1, 
%stage 1 lung function test
%stage 2 

For the COPD trial, an interim treatment selection was performed. 
Indacaterol is a drug administered to COPD patients.
The patients were randomized to four doses of indacaterol (\SI{75}{\micro\gram}, \SI{150}{\micro\gram}, \SI{300}{\micro\gram} and \SI{600}{\micro\gram}), to active controls, and to a placebo control.
Here, as in \citet{friede_adaptive_2020}, we ignore the active controls.
The primary outcome of the trial was the percentage of days of poor control over 26 weeks of the COPD patients.
Since the recruitment took only a short time, another outcome was required for treatment selection, in this case forced expiratory volume in 1 s (FEV1) at 15 days.
In the original study, the difference in FEV1 compared to placebo at 15 days for the different doses and corresponding confidence intervals were estimated.
From these, standardized effect sizes were estimated as approximately 0.68, 0.82, 0.95, and 0.91 for the four indacaterol doses \SI{75}{\micro\gram}, \SI{150}{\micro\gram}, \SI{300}{\micro\gram}, and \SI{600}{\micro\gram}, respectively.
For the final outcome days of poor control, also, from real data the standardized effect sizes were estimated.
These were approximately 0.13, 0.17, 0.23, and 0.20, for the four doses.
The approximate sample sizes per arm were 100 patients in stage 1 and 300 patients in stage 2.

In the analysis of \citet{friede_adaptive_2020}, the aim was to select two doses of indacaterol at the interim analysis that are passed to stage 2, resulting in a total sample size of $5 \cdot 100 + 3 \cdot 300 = 1,400$ patients. 
In the corresponding simulations, a positive correlation between early and final outcomes of $0.4$ was assumed.
Further, different settings with continuous early and continuous or binary final outcomes and different treatment selection strategies were compared.

% FIXME Question: 
% What does exactly specifies 'corr'?
% auf der individuen ebene

% What about these three papers?
%\cite{barnes_integrating_2010}
%\cite{donohue_oncedaily_2010}
%\cite{cuffe_when_2014}

\section{Adaptive Seamless Designs}
\label{sec:adaptive_seamless_designs}

% Is there a condensed version? In Friede et al (2020) it is only explained in detail?
% TODO: Tim
Traditionally, drug development is organized in four phases with individual clinical trials for the separate phases. Seamless designs combine elements from different development phases in a single trial, thereby offering the promise of speeding the development process. For instance, seamless phase II/III designs combine the learning about dose-response relationship or subgroup heterogeneity typical for phase II with the confirmation of treatment effects in phase III. If the data of the phase II part are used for confirmatory purposes, then this needs to be accounted for in the testing strategy to maintain the family-wise type I error rate in the strong sense. These types of designs are called adaptive seamless designs. A recent overview can be found in \citet{friede_adaptive_2020}.

We start by providing some more details on the study design considered (Sections \ref{ssec:two_stage_design_with_treatment_selection} and \ref{ssec:selection_rules}). Then, in Section \ref{ssec:clinical_scenario_evaluation_for_asd} we briefly introduce a framework for optimizing such designs, namely clinical scenario evaluation (CSE). %In Section \ref{ssec:simulating_asd} we make some brief comments on the simulation model.

\subsection{Two-stage design with treatment selection}
\label{ssec:two_stage_design_with_treatment_selection}
We consider a particular type of adaptive seamless designs, namely two-stage designs with treatment (or more specifically dose) selection. The trial starts in stage 1 with $K$ experimental treatments or doses of one experimental treatment which are compared to a common control. Interest is in assessing the treatment effects $\beta_k$ with $k=1,\dots, K$ comparing experimental treatment $k$ vs. control and testing the null hypotheses $H_{0,k}: \beta_k=0$ against the alternative $H_{1,k}: \beta_k>0$ (larger $\beta_k$ being better). For the purpose of treatment selection, test statistics $Z_{1,k}$ based on data from the first design stage are considered, with larger $Z_{1,k}$ indicating more beneficial treatments. These can use data on the primary outcome used for testing or any other outcome that is available at the time of the interim analysis. In practical applications, early outcomes are often used to inform interim decisions, since the primary endpoint might take a longer time to observe. In the interim analysis, treatments are selected for continuation into the second stage; the set of selected treatments is denoted by $\mathcal{S}$. The selection rules used here are described in Section \ref{ssec:selection_rules}. In the final analysis, the closed test principle is applied to account for the multiple hypotheses. The intersection hypotheses are tested by combining stagewise $p$-values for the respective intersection hypotheses using a prespecified combination function. Here we use the inverse normal combination function~\citep{lehmacher_adaptive_1999} which is given by
\begin{equation}
   w_1 \Phi^{-1}(1-p_{\mathcal{K},1}) + w_2 \Phi^{-1}(1-p_{\mathcal{K},2}) \,
\end{equation}
where $\Phi^{-1}(\cdot)$ denotes the quantile function of the standard normal distribution, $w_1$ and $w_2$ weights with $w_1^2+w_2^2 = 1$, and $p_{\mathcal{K},1}$ and $p_{\mathcal{K},2}$ stagewise $p$-values testing intersection hypothesis $H_{\mathcal{K}}$. The p-values for stages $1$ and $2$ are calculated on those patient recruited in stages $1$ and $2$, respectively. Note that one might not be able to calculate $p_{\mathcal{K},1}$ at the time of the interim analysis since some patients might still be under follow-up for the final endpoint.

\subsection{Selection rules}
\label{ssec:selection_rules}
In practice, the totality of the data would be considered by a data monitoring committee (DMC) to make a recommendation regarding the treatment selection. However, it is advisable for the sponsor to define a selection rule, although in comparison fairly simplistic, to provide some guidance for the DMC regarding the sponsor's preferences. Furthermore, formal selection rules are indispensable to conduct simulation studies, e.g. to explore sample size and power. In the following we give some examples of selection rules which have previously been considered and which we will feature below.

The treatment selection might consider statistics $Z_{1,k}$ for the final (primary) outcome or statistics $Z_{1,k}^{(E)}$ calculated based on some early outcome. We present the selection rules using $Z_{1,k}^{(E)}$, but of course these would be replaced by $Z_{1,k}$ if the final outcome was used for treatment selection. Again, larger statistics are considered better. 

\paragraph{$\kappa$-best Rule} Let $Z_{1,(1)}^{(E)}, \dots, Z_{1,(K)}^{(E)}$ denote the ordered statistics with $Z_{1,(1)}^{(E)}$ being the largest and therefore best test statistic. The $\kappa$-best rule then selects the treatments associated with the $\kappa$ largest values of the test statistic, %$Z_{1,k}^{(E)}$,
i.e. $Z_{1,(1)}^{(E)}, \dots, Z_{1,(\kappa)}^{(E)}$.

\paragraph{Epsilon Rule} To our knowledge this rule was first described by \citet{kelly_adaptive_2005} and has since then featured in a number of simulation studies including \citet{friede_comparison_2008} and \citet{friede_adaptive_2020}. All treatments $k$ with $Z_{1,k}^{(E)} \ge \max_i Z_{1,i}^{(E)} - \epsilon$ with $\epsilon \ge 0$ are selected. For $\epsilon=0$ this rule is equivalent to the $\kappa$-best rule with $\kappa=1$. 

\paragraph{Threshold Rule} With this rule all treatments $k$ are selected with $Z_{1,k}^{(E)} \ge \tau$, where $\tau$ is a given threshold. \\

We note that with the $\kappa$-best rule a fixed number of treatments is carried forward into the next stage whereas these are variable with the epsilon rule and the threshold rule.

\subsection{Clinical scenario evaluation for ASD}
\label{ssec:clinical_scenario_evaluation_for_asd}

\begin{figure}[ht]
\centering
\begin{tikzpicture}[auto,node distance=1cm,align=center,text width = 3cm, inner sep = 0.1cm, font=\sffamily \scriptsize, outer sep= 0cm]
  \node[draw, ellipse, minimum height=1cm, text width=2.5cm] (ClinicalScenarios) at (0,0) {
    \textbf{Clinical scenarios}\\
    Simulation Studies};
  \node[name=block, rectangle split, rectangle split parts=2, draw, above = of ClinicalScenarios] (UnknownPoints)
  { \begin{itemize}[leftmargin=.3cm, topsep=0cm,itemsep=0cm,partopsep=0cm, parsep=0cm]
      \item Treatment effect on early interim outcome
      \item Treatment effect on late disability outcome
    \end{itemize}
    \nodepart{second} Unknown
  };
  \node[name=block, rectangle split, rectangle split parts=2, draw, anchor = north east] (HistoricalDataPoints) at (UnknownPoints.north west)
  { \begin{itemize}[leftmargin=.3cm, topsep=0cm,itemsep=0cm,partopsep=0cm, parsep=0cm]
      \item Correlation between early interim and late disability outcomes
    \end{itemize}
    \nodepart{second} Historical data
  };
  %\node[draw, inner xsep=1em, inner ysep=1em, above left = of ClinicalScenarios, fit = (HistoricalDataPoints) (UnknownPoints)] (boxDiseaseSpecific) {};
  \node[draw, text width = 6.2cm, anchor = south west] (DiseaseSpecific) at (HistoricalDataPoints.north west) {\textbf{Disease-specific features}};
  %\node[draw, above right = of ClinicalScenarios] (DesignOptions) {Design options};
  \node[name=block, rectangle split, rectangle split parts=2, draw, anchor = north west] (ConstrainedPoints) at ([xshift = 4cm]HistoricalDataPoints.north east)
  { \begin{itemize}[leftmargin=.3cm, topsep=0cm,itemsep=0cm,partopsep=0cm, parsep=0cm]
      \item Number of test treatments
      \item Total subjects in trial
      \item Recruitment rate
    \end{itemize}
    \nodepart{second} Constrained by health care environment/infrastructure
  };
  \node[name=block, rectangle split, rectangle split parts=2, draw, anchor = north west] (CombinationPoints) at (ConstrainedPoints.north east)
  { Adaptive seamless design
    \begin{itemize}[leftmargin=.3cm, topsep=0cm,itemsep=0cm,partopsep=0cm, parsep=0cm]
      \item Time point of interim analysis
      \item Treatment selection rules at interim
    \end{itemize}
    \vspace*{0.5ex}
    Conventional phase 2/3
    \begin{itemize}[leftmargin=.3cm, topsep=0cm,itemsep=0cm,partopsep=0cm, parsep=0cm]
      \item Split of resources between phases
      \item End of phase futility criteria
    \end{itemize}
    \nodepart{second} Combination of options
  };
  \node[draw,text width = 6.2cm, anchor = south west] (DesignOptions) at (ConstrainedPoints.north west) {\textbf{Design options}};
  \node[draw, below = 0.5cm of ClinicalScenarios] (DesignPerformance) {
    \textbf{Design performance} \\ e.g.\ statistical power};
  \node[draw, right = of DesignPerformance, text width = 5cm, node distance = 2cm] (Evaluation) {
    \textbf{Clinical Scenario Evaluation}\\
    Design performance measures evaluated across a wide range of clinical scenarios};
  \draw[thick,->] (HistoricalDataPoints) -- (ClinicalScenarios);
  \draw[thick,->] (UnknownPoints) -- (ClinicalScenarios);
  \draw[thick,->] (CombinationPoints.223) -- (ClinicalScenarios);
  \draw[thick,->] (ConstrainedPoints) -- (ClinicalScenarios);
  \draw[thick,->] (ClinicalScenarios) -- (DesignPerformance);
  \draw[thick,->] (DesignPerformance) -- (Evaluation);
\end{tikzpicture}
%Friede et al (2010) DIJ, Figure 2 https://link.springer.com/content/pdf/10.1177/009286151004400507.pdf? https://link.springer.com/content/pdf/10.1177/009286151004400607.pdf
\caption{Application of the refined clinical scenario evaluation framework to adaptive seamless designs in progressive multiple sclerosis (Figure adapted from Figure~2 in \citet{friede_refinement_2010}).}
\label{fig:friede_refinement_2010_fig2}
\end{figure}
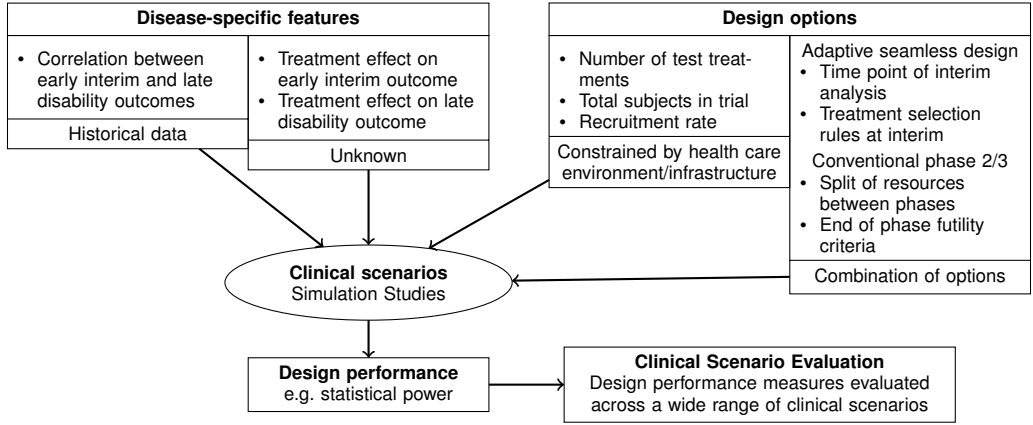

The clinical scenario evaluation framework was first proposed by \citet{benda_aspects_2010} and subsequently further developed by \citet{friede_refinement_2010}. 
Figure \ref{fig:friede_refinement_2010_fig2} provides a summary of this framework.
A clinical scenario evaluation is defined by the following three elements: (i) disease-specific features; (ii) design options; and (iii) design performance measures.
The disease-specific features describe distributional assumptions on the early and final outcomes including variances, correlations and treatment effect sizes.
Here, some can be estimated from data of previous clinical trials and some are simply unknown. Considering adaptive seamless designs, the design options include the time point of the interim analysis and the treatment selection rule, which can in principle both be chosen without any restrictions.
In contrast, some other design options such as the number of treatments and the total sample size are constrained by the environment and infrastructure available.
The performance measures (sometimes also referred to as metrics) in the context with adaptive seamless designs are, e.g., power, sample size distributions or duration of the study.
%\paragraph{Simulating ASD}
%\label{ssec:simulating_asd}

The evaluation of some performance measures requires Monte Carlo simulations, since no closed form solutions are available. 
The R-package \texttt{asd} by~\citet{parsons_package_2012} can be used for such simulations.
As it generates test statistics, following multivariate normal distributions, it is applicable to a wide range of outcomes, at least approximately, and generally more efficient in comparison to simulating individual participant data.
The simulation model was described in more detail by \citet{friede_adaptive_2020}. 
It features selection at interim based on early as well as final outcomes.

% Some corrections to the figure needed: e.g. design options (right panel)

\section{Bayesian Optimization}
\label{sec:bayesian_optimization}

Bayesian optimization (BO)~\citep{jones_taxonomy_2001}, also known as Model-based Optimization (MBO), is a state-of-the-art technique for expensive black-box optimization problems~\citep{shahriari_taking_2016}.
In comparison to other black-box optimization methods, like Genetic Algorithms or Simulated Annealing, BO is especially suitable when evaluating a configuration (e.g.\ running a simulation with certain parameters, here denoted by $\theta$) is very time-consuming.
In this situation it becomes infeasible to evaluate the black-box for thousands of configurations.

\subsection{General principle}
BO solves an optimization problem within a bounded search space $\Theta$:
\[
\theta^\ast := \operatorname{arg\,max}_{\theta \in \Theta} f(\theta),
\]
where $f(\theta)$ denotes the evaluation of the black-box with the input configuration $\theta$.
To reduce the number of evaluations on $f$ the key idea of BO is to only evaluate values of $\theta$ that are estimated to lead to a high value of $f(\theta)$.
The estimate is generated by a so called \emph{surrogate model}.
Typically, this is a regression model that predicts the outcome of $f$ based on previous evaluations of $f$.
First, an initial design of already evaluated configurations is needed.
Then, iteratively, the BO algorithm fits the surrogate on the previous evaluations, proposes a new configuration $\theta$ and evaluates it on $f$.
The steps are repeated until a budget is exhausted.

The proposal of a new configuration $\theta$ is obtained by maximizing a so-called acquisition function.
The acquisition function guides the search to promising new regions in the search space $\Theta$ by using the mean prediction $\hat{\mu}(\theta)$ as well as the uncertainty quantification $s^2(\theta)$ of the surrogate.
It balances between exploration of not yet evaluated regions in $\Theta$ and exploitation.
The acquisition function achieves exploration by assigning high values to areas where the surrogate predicts a high uncertainty.
Exploitation is achieved by assigning high values to areas of $\theta$ where the surrogate predicts a high outcome of $f$.
For deterministic functions the expected improvement~\citep{jones_efficient_1998} is arguably the most popular acquisition function.

The expected improvement as well as other acquisition functions are derived under the assumption that at each point $\theta \in \Theta$ the posterior of the function outcome follows a normal distribution.
The standard Gaussian process regression, that is often used as surrogate, models the posterior as a Gaussian with the parameters $\hat{\mu}$ and $s^2(\theta)$ under the assumption that $f$ is a realization of a Gaussian process.
In practice, this assumption often cannot be verified.
However, benchmarks show that the approach nevertheless works reliable for many practical problems~\citep{bischl_mlrmbo_2017,snoek_practical_2012}.

\subsection{Acquisition function for noisy outcomes}
Since this work focuses on optimization of a stochastic simulation, we assume that our non-deterministic optimization problem can be formulated as follows:
\begin{equation}
\theta^\ast := \operatorname{arg\,max}_{\theta \in \Theta} f(\theta) + \varepsilon, \, \text{ with } \, \varepsilon \sim N(0, \sigma^2) %\text{ i.i.d.}
\end{equation}
In this setting, the expected improvement is not feasible, as shown in \citet{huang_global_2006}.
They propose to use the augmented expected improvement instead.
Therefore, in a first step, we calculate the \emph{effective best solution}:
\begin{equation}
  \theta^{\ast\ast} := \operatorname{arg\,max}\limits_{\theta \in \mathcal{D}} \hat{\mu}(\theta) - c \cdot s(\theta),
\end{equation}
with $c$ as a tuning parameter that is usually set to 1 and $\mathcal{D}$ as the design that contains all previously evaluated values of $\theta$.
The effective best point is the pessimistic estimate of the best observed outcome so far.
In the final step, we calculate the augmented expected improvement~\citep{huang_global_2006} as follows:
\begin{multline}
  \label{eq:aei}
  \operatorname{AEI}(\theta) = \left( \hat{\mu}(\theta) - \hat{\mu}(\theta^{\ast\ast}) \right) \cdot  \Phi \biggl( \frac{\hat{\mu}(\theta) - \hat{\mu}(\theta^{\ast\ast})}{s(\theta)} \biggr) \\
   + s(\theta) \cdot \phi \biggl( \frac{\hat{\mu}(\theta) - \hat{\mu}(\theta^{\ast\ast})}{s(\theta)} \biggr) \cdot \biggl(1 - \underbrace{\frac{\sigma}{\sqrt{\sigma^2 + s^2(\theta)}}}_{\text{correction}}\biggr),
\end{multline}
where $\Phi$ and $\phi$ are the distribution and density function of the standard normal distribution, and $\sigma^2$ denotes the random error (nugget effect) in the Kriging model.
The formula is derived under the assumption of a Gaussian posterior at each point $\theta$.
The intuition in formula (\ref{eq:aei}) is that the first summand favors exploitation while the second summand favors exploration.
The correction term is necessary to avoid exploration in areas where the model uncertainty $s^2(\theta)$ equals the random error $\sigma^2$, because in this case further evaluations will not decrease the model uncertainty.

\subsection{Selection of the best point from noisy outcomes}
\label{ssec:best_point}
If we chose the configuration $\theta$ from all previously evaluated configurations according to the best outcome of $f$ as the optimization result, we would be overly optimistic.
It is likely that a single best outcome can be partially attributed to the random error $\sigma^2$.
Instead, we are interested in the optimum of the true posterior mean.
Therefore, we employ the surrogate estimate of the posterior mean for each evaluated configuration to cancel out the noise.
The configuration for which the surrogate estimates the best outcome is then returned as the optimization result $\hat{\theta}^\ast$:
\begin{equation}
  \hat{\theta}^{\ast} := \operatorname{arg\,max}\limits_{\theta \in \mathcal{D}} \hat{\mu}(\theta),
\end{equation}
%To obtain a final outcome $\nu^\ast = f(\theta^\ast)$ that is independent of the optimization process, $f(\theta^\ast)$ is calculated again.
Using the stochastic outcome $f(\hat{\theta}^\ast) $ observed during the optimization process would still potentially lead to an overly optimistic result.
Therefore, an independent calculation of $f(\hat{\theta}^\ast)$ should be conducted to obtain a fair estimate of the true value of the best outcome.

\section{Simulation Study}
\label{sec:simulation_study}

The goal of this simulation study is to investigate the potential benefit of BO for efficiently finding a clinical trial design that maximizes a given metric.
The analysis is motivated by the clinical example with COPD patients introduced in Section~\ref{sec:motivating_example}, with treatment selection in the first of two stages.
In the first stage five arms (defined by the control and four different administered doses) are considered, and in the second stage only a selected subset of arms is kept for further evaluation. 
The metric to be optimized is the statistical power, where a positive outcome means a trial arm with true effect is detected as significant effect.
Depending on the application the power can be calculated as the proportion of correctly rejecting any, all, or a subset of the elementary hypotheses~\citep{senn_power_2007}.

\subsection{Clinical trial design and parameters}
\label{subsec:design}
For the true effect sizes in the simulation, we consider four \emph{effect sets} whose values are inspired and partially based on the corresponding numbers of the motivating example, see Table~\ref{tab:table_effect_names}.
\begin{table}

\caption{\label{tab:table_effect_names}Effect sizes used for simulation in the first (early) and second (final) stage}
\centering
\begin{tabular}[t]{llrrrrr}
\toprule
\multicolumn{2}{c}{ } & \multicolumn{5}{c}{Treatment} \\
\cmidrule(l{3pt}r{3pt}){3-7}
Effect Set & Stage & 0 & 1 & 2 & 3 & 4\\
\midrule
 & early & 0 & 0.680 & 0.82 & 0.950 & 0.91\\

\multirow{-2}{*}{\raggedright\arraybackslash paper} & final & 0 & 0.130 & 0.17 & 0.230 & 0.20\\
\cmidrule{1-7}
 & early & 0 & 0.200 & 0.40 & 0.600 & 0.80\\

\multirow{-2}{*}{\raggedright\arraybackslash linear} & final & 0 & 0.050 & 0.10 & 0.150 & 0.20\\
\cmidrule{1-7}
 & early & 0 & 0.100 & 0.20 & 0.700 & 0.80\\

\multirow{-2}{*}{\raggedright\arraybackslash sigmoid} & final & 0 & 0.025 & 0.05 & 0.175 & 0.20\\
\cmidrule{1-7}
 & early & 0 & 0.680 & 0.82 & 0.950 & 0.91\\

\multirow{-2}{*}{\raggedright\arraybackslash paper2} & final & 0 & 0.260 & 0.34 & 0.460 & 0.40\\
\bottomrule
\end{tabular}
\end{table}

For the effect set \emph{paper} we use exactly the same numbers as in~\citet{friede_adaptive_2020}.
For the effect set \emph{paper2} the effect sizes for the second stage are doubled.
As further case we consider a \emph{linear} effect set with effect sizes linear increasing from 20\% to 80\% for the first stage and effect sizes divided by 4 for the second stage as well as a \emph{sigmoid} relationship, again with 4 times larger effect sizes for the first stage.
These sets reflect different realistic situations.

The strategies for treatment selection for the second stage in the two-phase trial design and their corresponding parameters that are used in the simulation study are given in Table~\ref{tab:selection_strategies} and Table~\ref{tab:search_space}, respectively.
The strategies follow the concept of adaptive seamless designs discussed in Section~\ref{sec:adaptive_seamless_designs}. 
The $\kappa$-best strategies select the 1, 2, 3 or 4 best strategies according to their respective test statistics in the first stage (with \emph{all} representing all $k=4$ treatments besides the control). 
The strategies \emph{eps} and \emph{thresh} represent strategies with a flexible number of treatments taken to the second stage. 
In Table~\ref{tab:search_space} the ranges for these two parameters and for the proportion of the total samples (ratio $r$) allocated to the second stage are given.
\begin{table}[h]
  \caption{Overview of selection strategies to determine which trial arms are kept for second stage}
  \label{tab:selection_strategies}
  \centering
  \begin{tabular}{ll}
  \hline
  Notation       & Procedure of selection strategy  \\
  \hline
  \emph{$\kappa$-best}    & Select $\kappa = 1, 2, 3$ or 4 (all) arms, respectively, according to maximal test statistic \\
  \emph{eps}     & Select all arms with test statistic less than $\epsilon$ (epsilon) below maximal test statistic \\
  \emph{thresh}  & Select all arms with test statistic larger than the value $\tau$ (threshold) \\
  \hline
  \end{tabular}
\end{table}
\begin{table}[h]
  \caption{Parameters for the selection strategies representing the search space of trial designs}
  \label{tab:search_space}
  \centering
  \begin{tabular}{ll}
  \hline
  Parameter                        & Range \\
  \hline
  Selection strategy               & \emph{\{1-best, 2-best, 3-best, all, eps, thresh\}} \\
  $r$ (ratio)                      & $(0,1)$ \\
  $\epsilon$ (only for \emph{eps}) & $[0,4]$ \\
  $\tau$ (only for \emph{thresh})  & $[0,10]$ \\
  \hline
  \end{tabular}
\end{table}

\subsection{Calculation of the statistical power}

The simulation of the statistical power of the trial design can be formulated as a black-box function as follows:
\begin{align}
  \label{eq:bbox}
  y & = f(n_{\text{stage1}}, n_{\text{stage2}}, \text{selection strategy}, \epsilon, \tau, \text{effect set}) \ ,
\end{align}
whereas $n_{\text{stage1}}$ and $n_{\text{stage2}}$ be the numbers of patients (samples) allocated to each treatment in stage 1 and stage 2 of the trial, respectively and \emph{effect set} remains fixed as we are interested in the optimal trial design for a given \emph{effect set}.
Our main optimization goal is to maximize the statistical power $y$ depending on the trial design. 

However, we are interested in keeping the total sample size $n_{\text{total}}$ constant in the overall trial for a fair comparison of the power of different designs.
Therefore, we introduce the ratio of patients (samples) per treatment in stage 1, compared to patients per treatment in both stages
\begin{align}
  \label{eq:r}
 r & = \frac{n_{\text{stage1}}}{n_{\text{stage1}}+n_{\text{stage2}}}
\end{align}
as a new parameter that will replace $n_{\text{stage1}}$ and $n_{\text{stage2}}$ in~(\ref{eq:bbox}). 

In the following we will explain how we derive the values of $n_{\text{stage1}}$ and $n_{\text{stage2}}$ in dependence of $r$ and $n_{\text{total}}$ so that we can maximize
\begin{equation}
  \label{eq:bbox2}
  y = \tilde{f}(r, \text{selection strategy}, \epsilon, \tau, \text{effect set}, n_{\text{total}})
\end{equation}
instead of~(\ref{eq:bbox}), whereas $n_{\text{total}}$ and \emph{effect set} remain fixed.

Let $k_1$ and $k_2$ denote the number of treatments in stage 1 and stage 2, respectively. 
Within one stage, the same number of patients are allocated to all different treatments. 
Then $n_{\text{total}}$ is given by
\begin{equation}
  \label{eq:ntreat}
  n_{\text{total}} = k_1 \cdot n_{\text{stage1}} + k_2 \cdot n_{\text{stage2}} \ .
\end{equation}
In our scenario (as described in Section \ref{subsec:design}) we have $k_1 = 5$ and $k_2 \in \{2, \ldots, 5\}$, where $k_2$ depends on the result of the treatment selection strategy in stage 1.
The control treatment is included in both stages.

Given the number of treatments $k_1$ and $k_2$ in the two stages and the total sample size $n_{\text{total}}$, optimizing the power with respect to $n_{\text{stage1}}$ and $n_{\text{stage2}}$ is equivalent to optimizing with respect to $r$, since
\begin{align}
  \label{eq:stagec}
  %\begin{split}
  n_{\text{stage1}} & = 
  n_{\text{stage1}}\cdot \frac{n_{\text{total}}}{k_1\cdot n_{\text{stage1}}+k_2\cdot n_{\text{stage2}}} = 
  n_{\text{total}} \cdot \frac{n_{\text{stage1}}}{k_1\cdot n_{\text{stage1}}+k_2\cdot n_{\text{stage2}}} \nonumber \\
  & = n_{\text{total}}\cdot
  \frac{r}{k_1\cdot
  r+k_2\cdot (1-r)} \ \text{and} \nonumber \\
  n_{\text{stage2}} & =
  n_{\text{total}}\cdot
  \frac{1-r}{k_1\cdot
  r+k_2\cdot (1-r)}.
  %\end{split}
\end{align}
%For $k_1 = k_2 = 5$, $c$ would be $n_{\text{total}} \cdot \frac{1}{5}$ because for each arm we can distribute $\frac{1}{5}$ of the treatments across stage 1 and stage 2.
These formulas give the numbers of patients (samples) per treatment arm in the two stages, if $k_2$ is known. 
This is only the case for selection strategies 1, 2, and 3.

For the selection strategies \emph{eps} and \emph{thresh}, $k_2$ depends on the outcome of stage 1.
%choice of the \emph{thresh} or \emph{eps} parameter.
In these cases, for given values of $n_{\text{stage1}}$ and $\epsilon$ or $\tau$, respectively, we can obtain an estimate $\hat{k}_2$ of $k_2$ with a so-called calibration step.
For a set of values for $n_{\text{stage1}}$ in the interval $\left[ \lceil 0.01 \cdot \frac{n_{\text{total}}}{k_1} \rceil, \lceil \frac{n_{\text{total}}}{k_1} \rceil \right]$, we simulate stage 1 of the trial multiple times.
We calculate $n_{\text{stage2}} = (1-r)/r\cdot n_{\text{stage1}}$ from formula (\ref{eq:r}) and $\hat{k}_2$ as the average of the resulting values of $k_2$ for all simulations with a fixed value for $n_{\text{stage1}}$.
Then we select the value of $n_{\text{stage1}}$ such that the corresponding total sample size is closest to $n_{\text{total}}$.
More precisely, we minimize
\begin{align}
  \label{eq:targettreat}
  \begin{split}
  h(n_{\text{stage1}}) &:= \left( \Big(k_1 \cdot n_{\text{stage1}} + \hat{k}_2 \cdot %\underbrace{
  \frac{1-r}{r} \cdot n_{\text{stage1}}
  %}_{=n_{\text{stage2}}}
  \Big) - n_{\text{total}} \right)^2
%  \hat{k}_2 &= f(n_{\text{stage1}}, 1, \text{effect set}, \text{Selection strategy}, \text{Epsilon}, \text{Threshold})_{k_2}
  \end{split}
\end{align}
with respect to $n_{\text{stage1}}$, where the estimate $\hat{k}_2$ is obtained as described above dependent on $n_{\text{stage1}}$. 
%$f$ is the simulator black-box from Equation~\ref{eq:bbox}, that gives us the estimated number of $\hat{k}_2$.
%This number is independent of the simulation of the second stage, so we can set $n_{\text{stage2}}$ to 1.
%The value of $\hat{k}_2$ that was observed with the minimal outcome of $h$ is then used to calculate $c$ in equation~\ref{eq:ntreatcc}.
The calibration step is necessary to guarantee that the total sample size $n_{\text{total}}$ is (on average) the same, also for the variable selection strategies \emph{eps} and \emph{thresh}.

The above procedure gives us the values for $n_{\text{stage1}}$ and $n_{\text{stage2}}$, allowing us in the following to find the optimal clinical trial design for a given \emph{effect set} and a fixed total sample size, by optimizing the parameter values for $r$, \emph{Selection strategy}, $\epsilon$ and $\tau$ through maximizing $\tilde f$ in~(\ref{eq:bbox2}).

\subsection{Optimization strategies}

In the simulation study we compare the ability of two optimization approaches for finding an optimal parameter configuration, on the one hand an exhaustive grid search on the parameter space and on the other hand BO (Bayesian optimization).
For different scenarios we compare the statistical power of the best found clinical trial design.
We also compare the runtime needed to determine the respective optimization results.
Next we introduce the two optimization strategies.

\paragraph{Grid and Grid small}
We conduct an \emph{exhaustive grid search} with a resolution of $l$ points per real-valued dimension and one point for each categorical dimension.
Referring to our application, for each selection strategy we evaluate a numerical grid with a resolution of $l$.
All 6 selection strategies have the numeric parameter $r$.
The strategies \emph{epsilon} and \emph{thresh} both have an additional second parameter, $\epsilon$ or $\tau$, respectively.
For $l=25$, we call this approach \textbf{Grid},
with a total of $4 \cdot 25 + 2 \cdot 25 \cdot 25 = 1350$ design configurations that are evaluated for one scenario.
We chose $l=25$ generously to obtain an estimate of the optimum that should be close to the true optimum.
Previous studies used half the resolution to assess the behavior of similar problems~\citep{friede_adaptive_2020}.
We define \textbf{Grid Small} to be a subset of this grid with $l=7$ points per dimension, resulting in $4 \cdot 7 + 2 \cdot 7 \cdot 7 = 126$.
This results in a number of evaluations that can be easily carried out in practice.

Both grids are evaluated with 20 stochastic replications to estimate the variance of the estimated power.

The best point per evaluated grid is determined by selecting the configuration with the best outcome, i.e.\ the concrete trial design with the largest power.
The corresponding value is generally too optimistic, as the random noise on the estimated value can contribute to the quality of the outcome.
To account for this optimism and report unbiased performance estimates, for each of the 20 replicate simulations, we identify the best configuration and consider the corresponding 19 outcomes of the other 19 replicate simulations, similar to a cross-validation framework in machine learning applications. 
%This eliminates the risk of only reporting outcomes that are optimal by chance.

\paragraph{BO (Model-based optimization) and BO Grid}
For applying BO, we first define how the surrogate model that predicts the outcome of $f$ for unknown values of $\theta$ (parameter configurations) is generated.
As regression model for the surrogate we choose Kriging (also called Gaussian process regression) and use the implementation in the R-package \texttt{DiceKriging}~\citep{roustant_dicekriging_2012}, because it is known to work well for fairly low dimensional search spaces.
We configure Kriging to apply the Matérn$\frac{5}{2}$ kernel with an estimated \emph{nugget effect} to account for the noisy response of $f$ and without scaling the input variables to $[0,1]$.
Note, that by doing so, we implicitly assume that $f$ is a realization of a Gaussian process and that each single outcome is a realization of a Gaussian.
However, as stated in Section~\ref{sec:bayesian_optimization}, we can expect the method to work well even if the assumption is slightly violated.

Kriging expects a numerical input, but our search space contains categorical parameters and is even hierarchical.
First, to transform the categorical parameters of the search space (see Table~\ref{tab:search_space}) into numerical values we use one-hot encoding.
Furthermore, our search space is hierarchical because the parameters $\epsilon$ and $\tau$ are only active for the respective selection strategies \emph{eps} and \emph{thresh}.
The hierarchical structure introduces inactive values for the configurations ($\theta$), i.e.\ if \emph{selection strategy} is set to \emph{eps} then the parameter $\tau$ of \emph{thresh} is inactive.
If a value is inactive, we set it to a value two times as high as the maximum in its active range (e.g.\ if $\tau$ is inactive it will be set to 20).
This ensures that within the original range the Kriging model is not or only minimally affected by the inactive values.
Note, that this trick only works because we did not scale the input variables to $[0,1]$ and because the estimated covariance of the Matérn$\frac{5}{2}$ kernel is approximately zero for distances greater than $4$.
With the above steps we ensure that the input for the Kriging is purely numeric.

As an acquisition function we choose the $\operatorname{AEI}$ as explained in Section~\ref{sec:bayesian_optimization}.
We start BO with an initial design of $16$ randomly sampled points, following an established rule-of-thumb of $4$ points per dimension, which is also well in line with the recommendation of using around $10\%$ of the budget for the initial design~\citep{bossek_initial_2020}.
We allow $100$ further iterations, summing up to a total budget of $116$ evaluated configurations of the simulator $\tilde{f}$.

At the end of the optimization procedure, the best configuration is chosen according to the best prediction obtained on the set of evaluated points as explained in Section~\ref{ssec:best_point}.
Again, overoptimism must be avoided. 
Thus, the best selected parameter configuration is then independently evaluated $20$ times again.

The whole BO procedure is repeated $20$ times to estimate the variance of the optimal solution found.

Additionally, we investigate how \emph{BO} performs when it is limited to return configurations that are within the grid introduced previously.
Therefore, we introduce \textbf{BO Grid}, which executes the optimization exactly as \emph{BO}, but the best configuration is mapped to the closest value on the grid. 
In other words, for \emph{BO Grid} we conduct the independent evaluation using the configuration within the grid that is closest to the optimal configuration that is proposed by \emph{BO}.
The comparison of the results of \emph{BO Grid} and \emph{Grid} will allow two possible conclusions:
First, if \emph{BO Grid} is worse than \emph{BO}, we can assume that the resolution of \emph{Grid} is not high enough to find an optimal configuration.
Second, if \emph{BO Grid} is better than \emph{Grid}, it shows that \emph{Grid} did not select the optimal configuration, due to noise.
Summing up, there are two factors that can prevent \emph{Grid} from finding the optimal configuration: lack of resolution and noise of the outcome.

\subsection{Implementation}

The algorithms used in this study are implemented in R.
For the Bayesian optimization, the R-package \texttt{mlrMBO}~\citep{bischl_mlrmbo_2017} is used.
For the simulation of clinical trials, the function \texttt{treatsel.sim} of the R-package \texttt{asd}~\citep{parsons_package_2012} is used, with parameter values given in Table~\ref{tab:par_implement}.
%The vector of treatment numbers for determining power \texttt{ptest} counts rejections of treat3 or 4 hypotheses. %???
The vector of treatment numbers \texttt{ptest} determines the hypotheses for which rejections are counted towards the power.
Here, the rejection of one or both of the hypotheses that treatment 3 or 4 has no effect against the control determines the power, as in~\citet{friede_adaptive_2020}.

Note that the parameter \texttt{nsim} denotes the internal simulation iterations of $f$. 
The 20 additional stochastic repetitions we conduct for the \emph{grid search} and \emph{BO} are additional replicates.

\begin{table}[h]
  \caption{Parameter choices for the R function \texttt{treatsel.sim}}
  \label{tab:par_implement}
  \centering
  \begin{tabular}{lll}
  \hline
  Meaning of parameter & Parameter & Value \\
  \hline
  Number of simulation iterations & \texttt{nsim} & $1000$ \\
  Correlation between early and final outcomes & \texttt{corr} & $0.4$ \\
  One-sided significance level & \texttt{level} & $0.025$ \\
  Vector of treatment numbers for determining power & \texttt{ptest} & $(3,4)$ \\
  \hline
  \end{tabular}
\end{table}

\subsection{Main results for comparison of algorithms}

In this chapter we present an overview of the results for the comparison of the four optimization strategies: exhaustive \emph{Grid} search, grid search with a reduced resolution (\emph{Grid Small}), \emph{BO} and \emph{BO Grid}.
We focus our comparison on independent, replicated evaluations of the configurations proposed by each optimizer.
The evaluations provide unbiased estimates of the performance ($y_{\text{valid}}$) of all optimizers.

For this study we consider 12 scenarios.
The scenarios are generated by combining the four different effect sets (see Table~\ref{tab:table_effect_names}) with the three different numbers for $n_{\text{total}}$.
Each optimizer is applied on each scenario with 20 stochastic repetitions.

\begin{figure}[tbh]
\centering
\includegraphics[width=\linewidth]{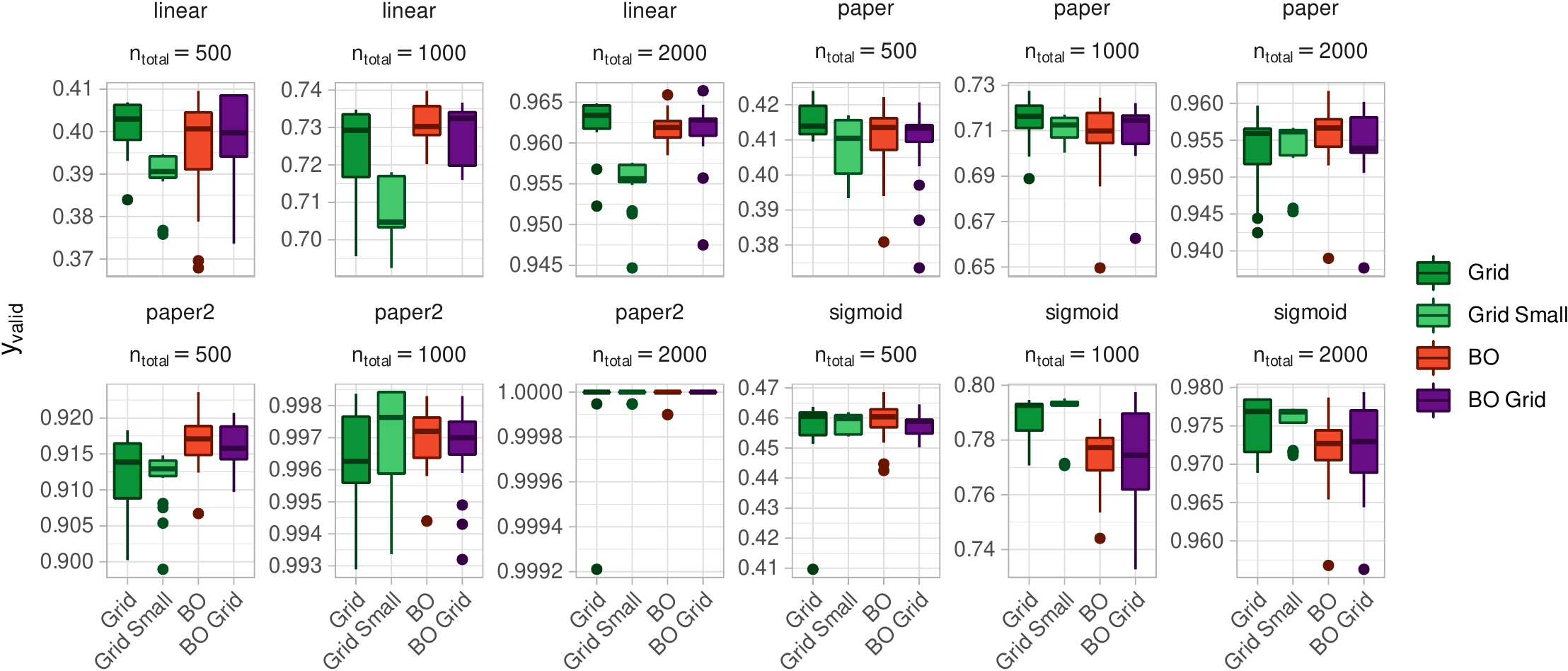}
\caption{%
  Box plots of performance (power) values on independent replicates of the best solutions for the four optimizers, the different effect sets, and different numbers of $n_{\text{total}}$ (500, 1000, 2000)
  }
\label{fig:plot_boxplot_valid_y}
\end{figure}

Figure~\ref{fig:plot_boxplot_valid_y} shows the estimated power values $y_{\text{valid}}$ for the best solutions in each scenario.
The plots show that \emph{BO} generally gives results that are comparable to those obtained with the (exhaustive) \emph{Grid} method across all scenarios. 
Only in some cases \emph{Grid} has a slight advantage (\emph{effect: linear}, $n_{\text{total}} = 2000$; \emph{effect: sigmoid}, $n_{\text{total}} = 1000$; \emph{effect: sigmoid}, $n_{\text{total}} = 2000$).
Below we will see that these scenarios have in common that the optimal configuration is close to the border of the search space. 
In other cases \emph{BO} yields slightly better results (\emph{effect: paper}, $n_{\text{total}} = 2000$; \emph{effect: paper2}, $n_{\text{total}} = 500$; \emph{effect: sigmoid}, $n_{\text{total}} = 500$).
Across all scenarios, \emph{BO} and \emph{Grid} perform similarly, although due to its evaluation budget, \emph{BO} is allowed only 116 evaluations as opposed to the 1350 evaluations for \emph{Grid}, when using a resolution of 25 points per dimension.
The difference in the number of evaluations is clearly reflected in the average runtime of the optimization methods given in Table~\ref{tab:table_time}.
Runtimes were measured on a single core of a compute node with 2 Intel Xeon E5-2697v2 (2.70 GHz) with 12 Cores each and 512 GB of RAM in a non-exclusive usage of the node.
\emph{BO} is approximately 20 times faster.
\begin{table}

\caption{\label{tab:table_time}Average runtime in hours, for evaluating one grid and one optimization run of MBO}
\centering
\resizebox{\linewidth}{!}{
\begin{tabular}[t]{lrrrrrrrrrrrrr}
\toprule
\multicolumn{1}{c}{ } & \multicolumn{3}{c}{linear} & \multicolumn{3}{c}{paper} & \multicolumn{3}{c}{paper2} & \multicolumn{3}{c}{sigmoid} & \multicolumn{1}{c}{ } \\
\cmidrule(l{3pt}r{3pt}){2-4} \cmidrule(l{3pt}r{3pt}){5-7} \cmidrule(l{3pt}r{3pt}){8-10} \cmidrule(l{3pt}r{3pt}){11-13}
 & 500 & 1000 & 2000 & 500 & 1000 & 2000 & 500 & 1000 & 2000 & 500 & 1000 & 2000 & \emph{evals}\\
\midrule
Grid & 51.2 & 55.3 & 56.3 & 56.7 & 58.7 & 63.0 & 55.1 & 59.2 & 61.6 & 50.7 & 51.3 & 56.8 & 1350\\
Grid Small & 3.8 & 4.1 & 4.1 & 4.1 & 4.2 & 4.6 & 4.1 & 4.3 & 4.4 & 3.7 & 3.8 & 4.1 & 126\\
BO & 3.5 & 3.3 & 3.7 & 3.3 & 3.6 & 3.7 & 3.5 & 3.3 & 3.2 & 3.3 & 2.9 & 3.6 & 116\\
\bottomrule
\end{tabular}}
\end{table}

For \emph{BO Grid}, only solutions on the grid that was evaluated by \emph{Grid} are considered in the evaluation. 
Comparing the results of \emph{BO Grid} to those of \emph{BO} we mainly see similar performances for both.
This indicates that the resolution of \emph{Grid} is fine enough to achieve similar performance to \emph{BO}.

Comparing \emph{BO Grid} to \emph{Grid}, we observe that for the scenarios \emph{effect: paper}, $n_{\text{total}} = 500$ and \emph{effect: paper}, $n_{\text{total}} = 1000$, \emph{BO Grid} achieves slightly better performance than \emph{Grid}.
This suggests that the \emph{Grid} optimization selects the best outcome "optimistically" from all observed outcomes and does not take into account the stochasticity of the problem, i.e. in this case the noise on the predicted power. 
Accordingly, we assume that for the mentioned scenarios the noise was so high, that \emph{Grid} was misguided by too optimistic outcomes when selecting the final configuration as optimization result.
In our evaluation, the independent validation $y_{\text{valid}}$ represents the unbiased performance of this final configuration, and the overoptimism becomes apparent.
In such scenarios \emph{BO} performs better, since the final configuration is determined based on the mean prediction of the surrogate of \emph{BO}, as explained in Section~\ref{ssec:best_point}.
Further below we will study on a single scenario if a reduced noise due to a higher number of simulations \texttt{nsim} decreases the effect of the overoptimistic selection of \emph{Grid}.

In cases in which \emph{BO} was not able to find a better solution then \emph{Grid}, also \emph{BO Grid} cannot yield better results as it just selects the configuration on the Grid that is closest to the configuration found by \emph{BO}.
Therefore, any increase in performance of \emph{BO Grid} compared to \emph{BO} is purely due to chance.

To compare \emph{BO} with a grid search that takes approximately the same time (see Table~\ref{tab:table_time}) and number of evaluations, we consider \emph{Grid Small}.
Here, a resolution of 7 points per dimension leads to 126 evaluations, close to 116 allowed evaluations of \emph{BO}, with only a small advantage for \emph{Grid Small}.
Despite the 10 more evaluations allowed, in the majority of cases \emph{BO} is superior to \emph{Grid Small}, see Figure~\ref{fig:plot_boxplot_valid_y}.
The advantage is particularly apparent for all scenarios with effect set \emph{effect: linear} and for scenario \emph{effect: paper2}, $n_{\text{total}} = 500$.
Only for the scenarios (\emph{effect: sigmoid}, $n_{\text{total}} = 1000$) and (\emph{effect: sigmoid}, $n_{\text{total}} = 2000$) \emph{Grid Small} has a notable advantage over \emph{BO} and \emph{Grid}.
In the second case, the absolute differences are negligible and all methods lead to a high power.
Also, for this case we will later show that the optimal configuration is close to the border of the search space.

As stated above, the high resolution of the exhaustive \emph{Grid} probably leads to the selection of slightly suboptimal configurations close to the optimum due to stochasticity.
These points are not included in the more coarse grid used by \emph{Grid Small}.
In other words, if a configuration close to the optimum is included in the grid used by \emph{Grid Small}, then it is more likely to select it with less overoptimism, as the number of candidates in the region is smaller.

Note that between the scenarios the absolute differences of the optimization results are of different practical relevance. For example, for scenario \emph{effect: sigmoid}, $n_{\text{total}} = 2000$ \emph{Grid} and \emph{Grid Small} are superior to \emph{BO}, but the difference is smaller than $0.5\%$.

%After having looked at the best obtained outcome on our problem scenarios and established the ability of \emph{BO} to find equally good configurations as the much more demanding exhaustive \emph{Grid}

\subsection{Detailed analysis of identified configurations}

\begin{figure}[htb]
\centering
\includegraphics[width=\linewidth]{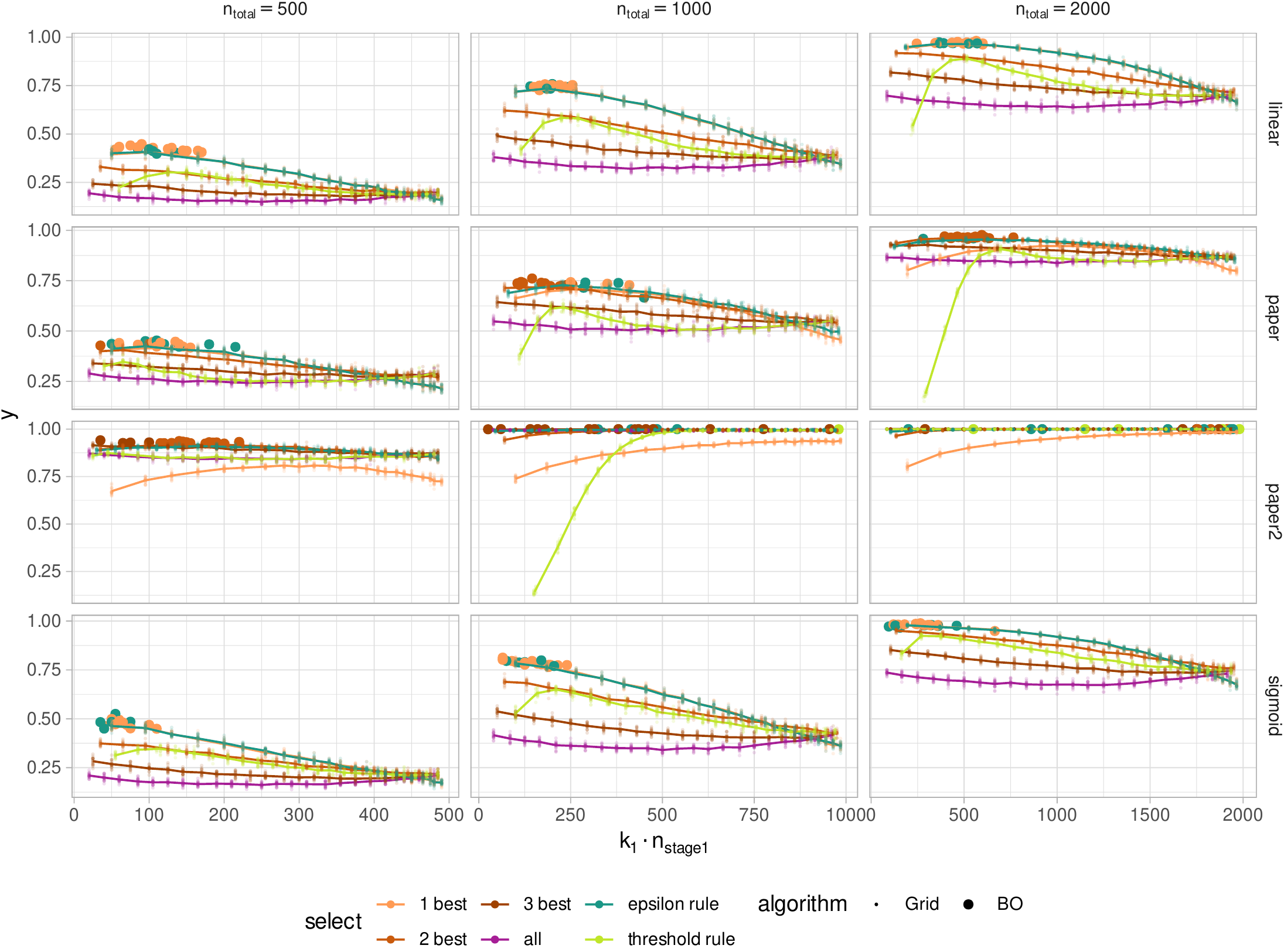}
\caption{%
  All evaluated configurations and their outcomes on the \emph{Grid}.
  For the epsilon and threshold rule only the curve for the $\epsilon$ and $\tau$ value with highest power ($y$) is displayed.
  For each configuration in the grid 20 %stochastic
  results are obtained.
  A line connects the mean outcomes for each configuration.
  In addition, the best configurations identified from \emph{BO} are displayed.
  }
\label{fig:plot_allbest}
\end{figure}

We now investigate the results in more detail.
We analyze the differences depending on the \emph{selection strategy} and the number of samples at stage 1, as determined by the \emph{stage ratio} ($r$).
Figure~\ref{fig:plot_allbest} shows the observed outcomes for all evaluations on the grid and the final configurations found by \emph{BO}.
For the selection strategies \emph{epsilon rule} and \emph{threshold rule} only curves representing the $\epsilon$ and $\tau$ values with the best outcome, respectively, are shown.
Note that for the effect sets \emph{linear} and \emph{sigmoid} the curves for selection strategy \emph{1 best} and \emph{eps} are nearly identical.
As intuitively anticipated, in almost all scenarios the worst results were obtained if all available samples are used in stage 1 and none in stage 2 (i.e.\ $r=1$).

In all scenarios, \emph{BO} identified the peak of all combined curves.
Within the 20 stochastic repetitions per scenario, \emph{BO} found different configurations, but all had similar high power.
The fact that no unique best setting was found across the repetitions can be attributed to the relatively high noise of the simulation and the relatively flat peaks, i.e.\ different configurations yield similarly good outcomes.
In cases in which multiple optimization runs result in different configurations it might be advisable to let a human expert choose the final configuration, also depending on the medical background.

The curves also show the potential of optimization for the different scenarios.
If the total sample size is high ($n_{\text{total}} = 1000$), then nearly all design configurations yield high statistical power and optimization is less important.
If the total sample size is too small ($n_{\text{total}} = 500$), then no configuration yields a practically usable design.
In this situation, a fast optimization helps to determine if a desired total sample size can actually yield sufficiently powerful designs.
Of special interest are the scenarios where the correct choice of the design configuration can make a difference between a design with acceptable power and a design with too low power, e.g. for the scenarios (\emph{effect: linear}, $n_{\text{total}} = 1000$; \emph{effect: linear}, $n_{\text{total}} = 2000$, \emph{effect: paper}, $n_{\text{total}} = 1000$; \emph{effect: sigmoid}, $n_{\text{total}} = 1000$).
Note, that the selection strategies \emph{thresh} and \emph{all} never selected powerful designs.

%TODO: Write more about the meaning of the results from a clinical point of view?

\begin{figure}[htb]
\centering
\includegraphics[width=\linewidth]{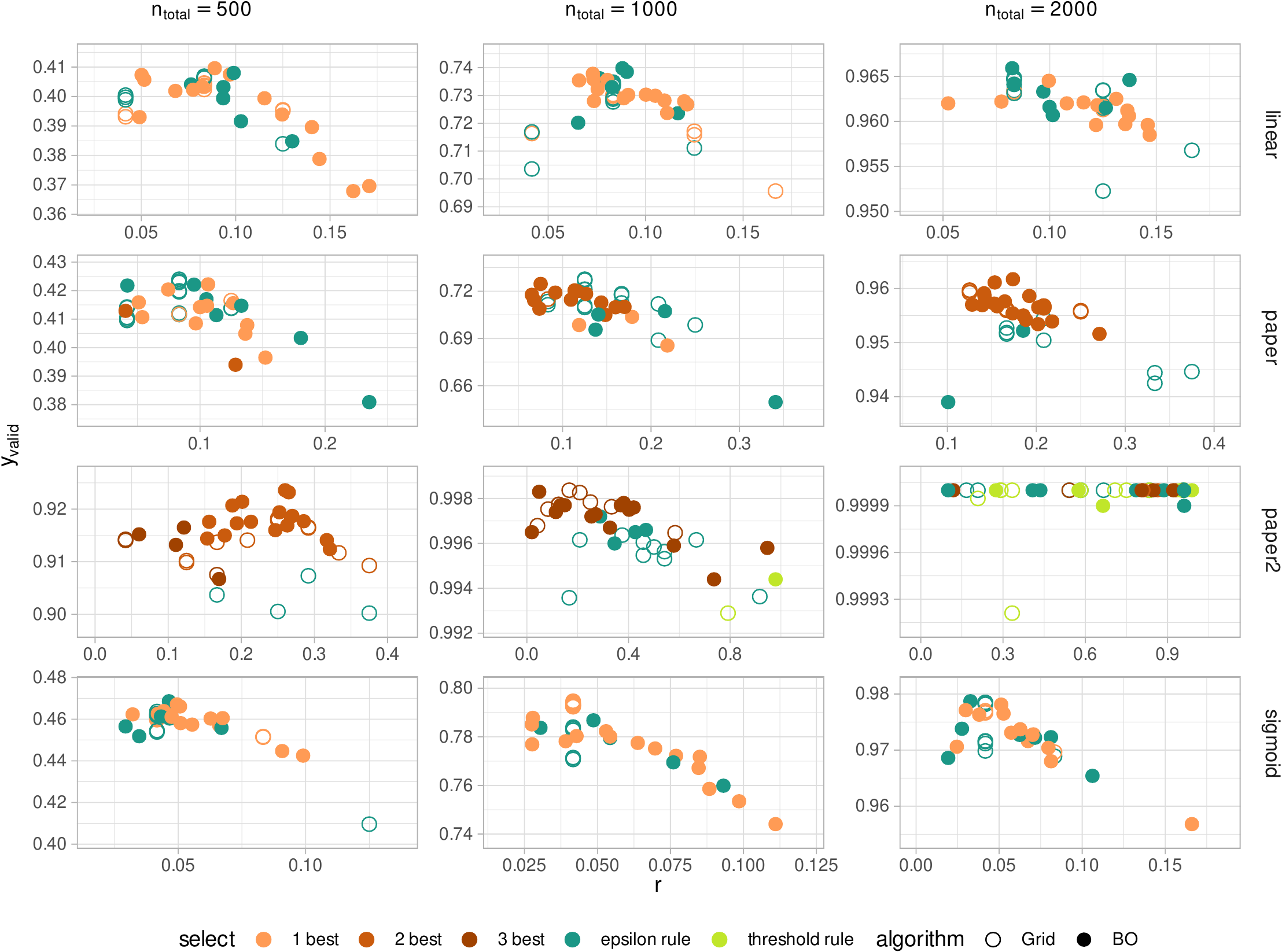}
\caption{Best configurations found with \emph{Grid} and \emph{BO} runs, i.e. corresponding $y_{\text{valid}}$ values for the ratio $r$ for the selected configurations. For the epsilon rule, $\epsilon$ and for the threshold rule $\tau$ are not displayed.}
\label{fig:plot_best_x}
\end{figure}

Figure~\ref{fig:plot_best_x} visualizes in detail which configurations were selected by \emph{BO} and \emph{Grid} and the corresponding values for $y_{\text{valid}}$ (result of the independent validation).
The optimal values for $\epsilon$ and $\theta$ are not shown.
For some scenarios most runs of \emph{Grid} returned the same optimal configurations (e.g.\ for \emph{effect: sigmoid}, $n_{\text{total}} = 1000$).
In such cases the signal-to-noise ratio was sufficiently high to reliably find the best configuration.
In some scenarios \emph{Grid} was superior to \emph{BO}, as could be seen in the previously shown Box plots in Figure~\ref{fig:plot_boxplot_valid_y}.
Furthermore, for all scenarios with \emph{effect: sigmoid} and the scenarios \emph{effect: linear}, $n_{\text{total}} = 500$ and \emph{effect: paper}, $n_{\text{total}} = 500$, small values of $r$ lead to an optimal outcome.
Here, \emph{BO} was not able to find these values close to the border of the search space.
A common remedy for this problem is to log-transform the values of~$r$.

In other cases \emph{Grid} found different configurations in different runs, while \emph{BO} lead to more stable results, also with higher power values (e.g.\ \emph{effect: linear}, $n_{\text{total}} = 1000$; \emph{effect: paper2}, $n_{\text{total}} = 500$; \emph{effect: sigmoid}, $n_{\text{total}} = 500$).
Here we can assume, that \emph{BO} successfully accounted for the noise.
A special case is scenario \emph{effect: paper2}, $n_{\text{total}} = 2000$:
Here nearly all found configurations yield perfect outcomes.

\subsection{Results for more simulation rounds}

To investigate the effect of the function noise variance ($\sigma^2$) on the optimization result we increase the simulation iterations from $n_\text{sim} = 1000$ to $n_\text{sim} = 5000$, restricting to the scenario \emph{effect: paper}, $n_{\text{total}} = 2000$.
% This is expected to increase the runtime by a factor of 5. % obvious, also not of interest here
Additionally, we run the algorithms 100 times instead of the previously used 20, in order to obtain more precise estimates for the power values in the independent validation.

We expect that increasing $n_\text{sim}$ decreases the variance of the power values of the identified design configurations.
For the exhaustive \emph{Grid}, we expect that the decreased noise leads to fewer selections of overly optimistic outcomes.
\emph{BO} should profit as well from a less noisy objective function since the predictions of the surrogate function will be more precise.

\begin{figure}[htb]
\centering
  \begin{minipage}{0.62\textwidth}
    \centering
    \includegraphics[width=\linewidth]{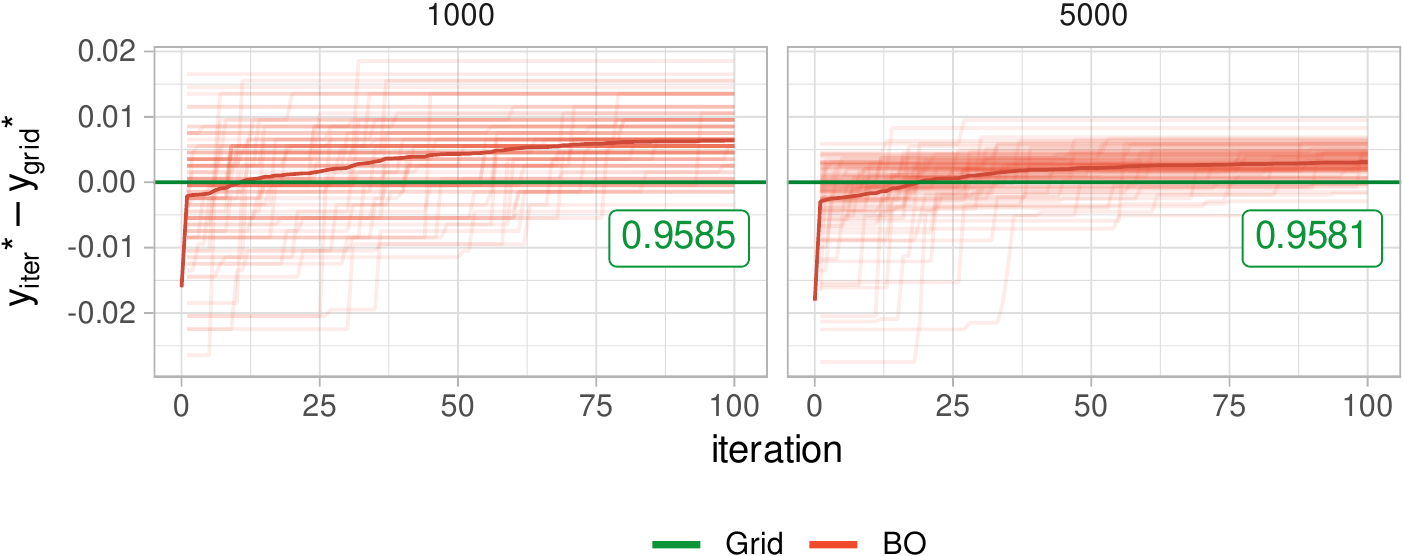}%
    \captionof{figure}{Optimization progress for \emph{BO}: Best outcome found so far for each iteration, in comparison to the best outcome found in the grid search, for effect set \emph{paper} and $n_{\text{total}} = 2000$}
    \label{fig:plot_opt_path_5000}
  \end{minipage}\hspace{0.02\textwidth}%
  \begin{minipage}{0.35\textwidth}
    \centering
    \includegraphics[width=\linewidth]{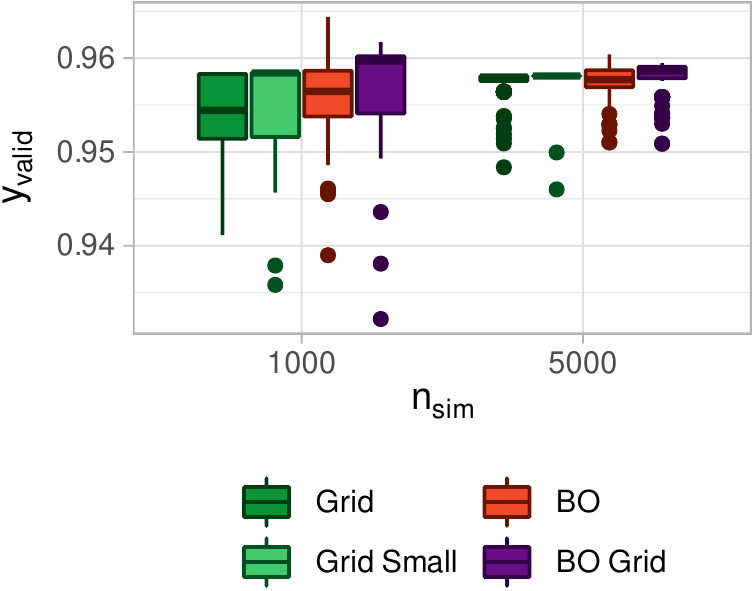}%
    \captionof{figure}{Validated performance of all optimization strategies on effect set \emph{paper} and $n_{\text{total}} = 2000$ with 100 stochastic repetitions and different values for \texttt{nsim}}
    \label{fig:plot_boxplot_valid_y_5000}
  \end{minipage}
\end{figure}

Figure~\ref{fig:plot_opt_path_5000} shows that indeed the lower noise level leads to more stable simulation runs, meaning that single optimization runs lead to results closer to the mean result.
Both curves indicate, that more iterations of \emph{BO} would not yield substantially better outcomes as no substantial improvements are observed after 80 iterations
It has to be noted that this plot reports the best observed outcome so far and not the one that is chosen as final result.
%So on the left side of Figure~\ref{fig:plot_opt_path_5000}, we see more curves that observed overoptimistic outcomes in comparison the the right side.
%Ideally, the choice of the final point should not be influenced by these overoptimistic observations for \emph{BO} as explained in Section~\ref{ssec:best_point}, as the final point is determined by the surrogates mean prediction.
Despite the differences of the optimization paths, the median validation error reported in Figure~\ref{fig:plot_boxplot_valid_y_5000} is nearly identical for \emph{BO} for the low and the high value of $n_\text{sim}$.
Therefore, we can assume that "bias correction" of \emph{BO} as explained in Section~\ref{ssec:best_point} seems to be effective.
For all methods the IQR displayed in the Box plots is much smaller for $n_\text{sim} = 5000$, indicating that the solutions that are found vary less between the 100 stochastic replications.
Also, the absolute differences between the methods are even smaller now.
This demonstrates that for the selected scenario all methods can find the optimal configuration if the results are not noisy.
However, an increased value of $n_\text{sim}$ also increases the runtime linearly, which conflicts with the goal to find the optimal configuration quickly.
In summary, \emph{BO} can find configurations with a small number of required simulations (higher noise) that are nearly as good as the optimal solution which is identified with a large number of simulations.

% This is the body text. Please note that cross-references in the body text should be shown as follows:
% (Miller, 1900), (Miller and Baker, 1900) or if three or more authors (Miller {\it{et al}}., 1900)
% \vspace*{12pt}

% \noindent Bullet lists are not allowed. Always use (i), (ii), etc.
% \vspace*{12pt}

% \noindent Sentences should never start with a symbol.
% \vspace*{12pt}

% \noindent Names of software packages and website addresses should be written in {\texttt{Courier new, i.e. Stata, the R package
% MASS, http://www.biometrical-journal.com.}}

\section{Discussion}
\label{sec:discussion}

% subsection Summary
Bayesian optimization (BO) is a state-of-the art optimizer that can be used for parameter optimization in a wide range of applications.
Here, we propose, for the first time, to optimize the power of adaptive seamless designs using BO.
In clinical trials, determining the test methods and sample sizes is a key step in the planning process.
For our optimization we chose to maximize the test power, given a set of treatments with corresponding effect sizes and a fixed number of samples.
%FIXME: TF: Sollten wir hier nochmal explizit auf alternative objectives wie study duration eingehen?
If determining the test power involves intensive simulations, trying out different treatment design configurations can become a laborious task.
Either large computational resources are required for an exhaustive investigation of all the different configurations, or a tedious manual trial-and-error process is required.

In this paper we analyzed if BO can quickly find an adaptive seamless design with high power, more efficiently than an exhaustive grid search.
For various sets of treatment effect sizes and total sample sizes, BO was able to find competitive designs in a fraction of the time.
In contrast to grid search, BO was able to account for stochasticity in the metric that should be optimized.
%In addition to results in other research fields, our results underline that BO is a powerful method for finding optimal configurations for problems that fit the black-box definition.
Therefore, the application of BO to clinical trial design selection in general seems promising. 
The same concept could be applied for subgroup selection in adaptive enrichment designs, which make use of interim data in selecting the target population for the remainder of the trial~\citep{burnett_adaptive_2020}.

% subsection Outlook
Only in cases in which the optimal values of one parameter are close to the lower border, we sometimes observed poor performance of BO.
Here, a possible improvement can be to apply a log transformation to such parameters. % in the future.
We chose to maximize the power for a fixed number of samples.
Other aspects like study duration or the cost of a trial in general could be of interest either as an additional constraint or an additional outcome.
Also the number of samples can be included as an additional outcome instead of being fixed.
Multiple outcomes that are subject to optimization give a (possibly restricted) multi-objective optimization problem.
The result of which is a set of points (a Pareto front) that maximizes the statistical power and e.g. minimizes the total sample size at the same time.
The user could then see for which sample size the power is above a certain threshold (e.g. $95\%$).
Another way to speed up the optimization would be to iteratively increase the number of simulation rounds of the trial design evaluation. 
If few simulations already indicate a bad outcome for a configuration then additional simulation rounds are not necessary.
Such a procedure can be guided by statistical testing and is already common in machine learning under the terms \emph{racing} or \emph{multi-fidelity optimization}.

% Assurance: \cite{stallard_optimal_2009} in diskussion abhandeln
% could be also applied for subgroup selection?
% https://onlinelibrary.wiley.com/doi/full/10.1002/sim.8797

In this work we assumed the effect sizes to be fixed. % under the null hypothesis.
In~\citet{stallard_optimal_2009} the authors propose a Bayesian approach, where the effect sizes follow a prior, to calculate the expected trial performance.
This requires a more expensive simulation making the optimization of a clinical trial design according to such metric an excellent candidate for BO.

% IDEA: Restricting BO to only evaluate configurations on the grid to avoid evaluating identical / very close configurations.

% \begin{table}[htb]
% \begin{center}
% \caption{The caption of a table.}
% \begin{tabular}{lll}
% \hline
% Description 1 & Description 2 & Description 3\\
% \hline
% Row 1, Col 1 & Row 1, Col 2 & Row 1, Col 3\\
% Row 2, Col 1 & Row 2, Col 2 & Row 2, Col 3\\
% \hline
% \end{tabular}
% \end{center}
% \end{table}
% \begin{equation}
% \left({\theta^{0}_{i}}\atop{\theta^{1}_{i}}\right) \sim N(\theta,\Sigma),\quad {\mathrm{with}}\ 
% {{\theta}} = \left({\theta_{0}}\atop{\theta_{1}}\right)\ {\mathrm{and}}\ \Sigma =
% \left(\begin{array}{cc}
% \sigma^{2}_{0} & \rho\sigma_{0}\sigma_{1}\\
% \rho\sigma_{0}\sigma_{2} & \sigma^{2}_{1}
% \end{array}\right).
% \end{equation}

%\noindent This is the body text. Only number equations which are referred to in the text body. If equations are numbered, these should be numbered continuously throughout the text. Not section wise! Please carefully follow the rules for mathematical expressions in the ``Instructions to Authors''.

\section*{Acknowledgments}
This work was partly supported by Deutsche Forschungsgemeinschaft (DFG) within the Collaborative Research Center SFB 876, A3.

%\section*{Appendix {\it(please insert here, if applicable)}}

%\subsection*{A.1.\enspace Second level heading}

%Please insert appendices before the references.

\bibliographystyle{unsrtnat}
\bibliography{literature,literature_zotero}

\begin{thebibliography}{31}
\providecommand{\natexlab}[1]{#1}
\providecommand{\url}[1]{\texttt{#1}}
\expandafter\ifx\csname urlstyle\endcsname\relax
  \providecommand{\doi}[1]{doi: #1}\else
  \providecommand{\doi}{doi: \begingroup \urlstyle{rm}\Url}\fi

\bibitem[Campbell et~al.(2000)Campbell, Fitzpatrick, Haines, Kinmonth,
  Sandercock, Spiegelhalter, and Tyrer]{campbell_framework_2000}
Michelle Campbell, Ray Fitzpatrick, Andrew Haines, Ann~Louise Kinmonth, Peter
  Sandercock, David Spiegelhalter, and Peter Tyrer.
\newblock Framework for design and evaluation of complex interventions to
  improve health.
\newblock \emph{BMJ}, 321\penalty0 (7262):\penalty0 694--696, September 2000.
\newblock ISSN 0959-8138, 1468-5833.
\newblock \doi{10.1136/bmj.321.7262.694}.

\bibitem[Sheiner(1997)]{sheiner_learning_1997}
Lewis~B. Sheiner.
\newblock Learning versus confirming in clinical drug development.
\newblock \emph{Clinical Pharmacology \& Therapeutics}, 61\penalty0
  (3):\penalty0 275--291, 1997.
\newblock ISSN 1532-6535.
\newblock \doi{10.1016/S0009-9236(97)90160-0}.

\bibitem[Bogin(2020)]{bogin_master_2020}
Vladimir Bogin.
\newblock Master protocols: New directions in drug discovery.
\newblock \emph{Contemporary Clinical Trials Communications}, 18:\penalty0
  1--5, June 2020.
\newblock ISSN 2451-8654.
\newblock \doi{10.1016/j.conctc.2020.100568}.

\bibitem[Hanneman(2008)]{hanneman_design_2008}
Sandra Hanneman.
\newblock Design, analysis, and interpretation of method-comparison studies.
\newblock \emph{AACN Advanced Critical Care}, 19\penalty0 (2):\penalty0
  223--234, April 2008.
\newblock ISSN 1559-7768.
\newblock \doi{10.1097/01.AACN.0000318125.41512.a3}.

\bibitem[Boulesteix et~al.(2013)Boulesteix, Lauer, and
  Eugster]{boulesteix_plea_2013}
Anne-Laure Boulesteix, Sabine Lauer, and Manuel J.~A. Eugster.
\newblock A plea for neutral comparison studies in computational sciences.
\newblock \emph{PLOS ONE}, 8\penalty0 (4):\penalty0 1--11, April 2013.
\newblock ISSN 1932-6203.
\newblock \doi{10.1371/journal.pone.0061562}.

\bibitem[Benda et~al.(2010)Benda, Branson, Maurer, and
  Friede]{benda_aspects_2010}
Norbert Benda, Michael Branson, Willi Maurer, and Tim Friede.
\newblock Aspects of modernizing drug development using clinical scenario
  planning and evaluation.
\newblock \emph{Drug Information Journal}, 44\penalty0 (3):\penalty0 299--315,
  May 2010.
\newblock ISSN 0092-8615.
\newblock \doi{10.1177/009286151004400312}.

\bibitem[Friede et~al.(2010)Friede, Nicholas, Stallard, Todd, Parsons,
  {Vald{\'e}s-M{\'a}rquez}, and Chataway]{friede_refinement_2010}
Tim Friede, Richard Nicholas, Nigel Stallard, Susan Todd, Nicholas Parsons,
  Elsa {Vald{\'e}s-M{\'a}rquez}, and Jeremy Chataway.
\newblock Refinement of the clinical scenario evaluation framework for
  assessment of competing development strategies with an application to
  multiple sclerosis.
\newblock \emph{Drug Information Journal}, 44\penalty0 (6):\penalty0 713--718,
  November 2010.
\newblock ISSN 0092-8615.
\newblock \doi{10.1177/009286151004400607}.

\bibitem[Jones(2001)]{jones_taxonomy_2001}
Donald~R. Jones.
\newblock A taxonomy of global optimization methods based on response surfaces.
\newblock \emph{Journal of Global Optimization}, 21\penalty0 (4):\penalty0
  345--383, 2001.
\newblock ISSN 0925-5001, 1573-2916.
\newblock \doi{10.1023/A:1012771025575}.

\bibitem[Snoek et~al.(2012)Snoek, Larochelle, and Adams]{snoek_practical_2012}
Jasper Snoek, Hugo Larochelle, and Ryan~P Adams.
\newblock Practical bayesian optimization of machine learning algorithms.
\newblock In F.~Pereira, C.~J.~C. Burges, L.~Bottou, and K.~Q. Weinberger,
  editors, \emph{Advances in Neural Information Processing Systems 25}, pages
  2951--2959. Curran Associates, Inc., 2012.

\bibitem[Hutter et~al.(2011)Hutter, Hoos, and
  {Leyton-Brown}]{hutter_sequential_2011}
Frank Hutter, Holger~H. Hoos, and Kevin {Leyton-Brown}.
\newblock Sequential model-based optimization for general algorithm
  configuration.
\newblock In Carlos A.~Coello Coello, editor, \emph{Learning and Intelligent
  Optimization}, number 6683 in Lecture Notes in Computer Science, pages
  507--523. Springer Berlin Heidelberg, January 2011.
\newblock ISBN 978-3-642-25565-6.
\newblock \doi{10.1007/978-3-642-25566-3_40}.

\bibitem[Balandat et~al.(2020)Balandat, Karrer, Jiang, Daulton, Letham, Wilson,
  and Bakshy]{balandat_botorch_2020}
Maximilian Balandat, Brian Karrer, Daniel~R. Jiang, Samuel Daulton, Benjamin
  Letham, Andrew~Gordon Wilson, and Eytan Bakshy.
\newblock Botorch: Programmable bayesian optimization in pytorch.
\newblock \emph{arXiv:1910.06403 [cs, math, stat]}, pages 1--34, June 2020.

\bibitem[Bischl et~al.(2017)Bischl, Richter, Bossek, Horn, Thomas, and
  Lang]{bischl_mlrmbo_2017}
Bernd Bischl, Jakob Richter, Jakob Bossek, Daniel Horn, Janek Thomas, and
  Michel Lang.
\newblock Mlrmbo: A modular framework for model-based optimization of expensive
  black-box functions.
\newblock \emph{arXiv:1703.03373 [stat.ML]}, pages 1--26, March 2017.

\bibitem[Wozniak et~al.(2018)Wozniak, Jain, Balaprakash, Ozik, Collier, Bauer,
  Xia, Brettin, Stevens, {Mohd-Yusof}, Cardona, Essen, and
  Baughman]{wozniak_candle_2018}
Justin~M. Wozniak, Rajeev Jain, Prasanna Balaprakash, Jonathan Ozik,
  Nicholson~T. Collier, John Bauer, Fangfang Xia, Thomas Brettin, Rick Stevens,
  Jamaludin {Mohd-Yusof}, Cristina~Garcia Cardona, Brian~Van Essen, and Matthew
  Baughman.
\newblock Candle/supervisor: A workflow framework for machine learning applied
  to cancer research.
\newblock \emph{BMC Bioinformatics}, 19\penalty0 (18):\penalty0 491, December
  2018.
\newblock ISSN 1471-2105.
\newblock \doi{10.1186/s12859-018-2508-4}.

\bibitem[Richter et~al.(2019)Richter, Madjar, and
  Rahnenf{\"u}hrer]{richter_modelbased_2019}
Jakob Richter, Katrin Madjar, and J{\"o}rg Rahnenf{\"u}hrer.
\newblock Model-based optimization of subgroup weights for survival analysis.
\newblock \emph{Bioinformatics}, 35\penalty0 (14):\penalty0 i484--i491, July
  2019.
\newblock ISSN 1367-4803.
\newblock \doi{10.1093/bioinformatics/btz361}.

\bibitem[Browaeys et~al.(2020)Browaeys, Saelens, and
  Saeys]{browaeys_nichenet_2020}
Robin Browaeys, Wouter Saelens, and Yvan Saeys.
\newblock Nichenet: Modeling intercellular communication by linking ligands to
  target genes.
\newblock \emph{Nature Methods}, 17\penalty0 (2):\penalty0 159--162, February
  2020.
\newblock ISSN 1548-7105.
\newblock \doi{10.1038/s41592-019-0667-5}.

\bibitem[Horn et~al.(2015)Horn, Wagner, Biermann, Weihs, and
  Bischl]{horn_modelbased_2015}
Daniel Horn, Tobias Wagner, Dirk Biermann, Claus Weihs, and Bernd Bischl.
\newblock Model-based multi-objective optimization: Taxonomy, multi-point
  proposal, toolbox and benchmark.
\newblock In \emph{International Conference on Evolutionary Multi-Criterion
  Optimization}, pages 64--78. Springer, 2015.

\bibitem[Wilson et~al.(2021)Wilson, Hooper, Brown, Farrin, and
  Walwyn]{wilson_efficient_2021}
Duncan~T Wilson, Richard Hooper, Julia Brown, Amanda~J Farrin, and Rebecca~EA
  Walwyn.
\newblock Efficient and flexible simulation-based sample size determination for
  clinical trials with multiple design parameters.
\newblock \emph{Statistical Methods in Medical Research}, 30\penalty0
  (3):\penalty0 799--815, March 2021.
\newblock ISSN 0962-2802, 1477-0334.
\newblock \doi{10.1177/0962280220975790}.

\bibitem[Barnes et~al.(2010)Barnes, Pocock, Magnussen, Iqbal, Kramer, Higgins,
  and Lawrence]{barnes_integrating_2010}
Peter~J. Barnes, Stuart~J. Pocock, Helgo Magnussen, Amir Iqbal, Benjamin
  Kramer, Mark Higgins, and David Lawrence.
\newblock Integrating indacaterol dose selection in a clinical study in copd
  using an adaptive seamless design.
\newblock \emph{Pulmonary Pharmacology \& Therapeutics}, 23\penalty0
  (3):\penalty0 165--171, June 2010.
\newblock ISSN 10945539.
\newblock \doi{10.1016/j.pupt.2010.01.003}.

\bibitem[Parsons et~al.(2012)Parsons, Friede, Todd, Marquez, Chataway,
  Nicholas, and Stallard]{parsons_package_2012}
Nick Parsons, Tim Friede, Susan Todd, Elsa~Valdes Marquez, Jeremy Chataway,
  Richard Nicholas, and Nigel Stallard.
\newblock An r package for implementing simulations for seamless phase ii/iii
  clinical trials using early outcomes for treatment selection.
\newblock \emph{Computational Statistics \& Data Analysis}, 56\penalty0
  (5):\penalty0 1150--1160, May 2012.
\newblock ISSN 0167-9473.
\newblock \doi{10.1016/j.csda.2010.10.027}.

\bibitem[Friede et~al.(2020)Friede, Stallard, and
  Parsons]{friede_adaptive_2020}
Tim Friede, Nigel Stallard, and Nicholas Parsons.
\newblock Adaptive seamless clinical trials using early outcomes for treatment
  or subgroup selection: Methods, simulation model and their implementation in
  r.
\newblock \emph{Biometrical Journal}, \penalty0 (62):\penalty0 1264--1283,
  March 2020.
\newblock ISSN 0323-3847, 1521-4036.
\newblock \doi{10.1002/bimj.201900020}.

\bibitem[Lehmacher and Wassmer(1999)]{lehmacher_adaptive_1999}
Walter Lehmacher and Gernot Wassmer.
\newblock Adaptive sample size calculations in group sequential trials.
\newblock \emph{Biometrics}, 55\penalty0 (4):\penalty0 1286--1290, December
  1999.
\newblock ISSN 0006341X.
\newblock \doi{10.1111/j.0006-341X.1999.01286.x}.

\bibitem[Kelly et~al.(2005)Kelly, Stallard, and Todd]{kelly_adaptive_2005}
Patrick~J. Kelly, Nigel Stallard, and Susan Todd.
\newblock An adaptive group sequential design for phase ii/iii clinical trials
  that select a single treatment from several.
\newblock \emph{Journal of Biopharmaceutical Statistics}, 15\penalty0
  (4):\penalty0 641--658, July 2005.
\newblock ISSN 1054-3406, 1520-5711.
\newblock \doi{10.1081/BIP-200062857}.

\bibitem[Friede and Stallard(2008)]{friede_comparison_2008}
Tim Friede and Nigel Stallard.
\newblock A comparison of methods for adaptive treatment selection.
\newblock \emph{Biometrical Journal}, 50\penalty0 (5):\penalty0 767--781,
  October 2008.
\newblock ISSN 03233847, 15214036.
\newblock \doi{10.1002/bimj.200710453}.

\bibitem[Shahriari et~al.(2016)Shahriari, Swersky, Wang, Adams, and
  de~Freitas]{shahriari_taking_2016}
B.~Shahriari, K.~Swersky, Z.~Wang, R.~P. Adams, and N.~de~Freitas.
\newblock Taking the human out of the loop: A review of bayesian optimization.
\newblock \emph{Proceedings of the IEEE}, 104\penalty0 (1):\penalty0 148--175,
  January 2016.
\newblock ISSN 0018-9219.
\newblock \doi{10.1109/JPROC.2015.2494218}.

\bibitem[Jones et~al.(1998)Jones, Schonlau, and Welch]{jones_efficient_1998}
Donald~R. Jones, Matthias Schonlau, and William~J. Welch.
\newblock Efficient global optimization of expensive black-box functions.
\newblock \emph{Journal of Global Optimization}, 13\penalty0 (4):\penalty0
  455--492, December 1998.
\newblock ISSN 0925-5001, 1573-2916.
\newblock \doi{10.1023/A:1008306431147}.

\bibitem[Huang et~al.(2006)Huang, Allen, Notz, and Zeng]{huang_global_2006}
D.~Huang, T.~T. Allen, W.~I. Notz, and N.~Zeng.
\newblock Global optimization of stochastic black-box systems via sequential
  kriging meta-models.
\newblock \emph{Journal of Global Optimization}, 34\penalty0 (3):\penalty0
  441--466, March 2006.
\newblock ISSN 0925-5001, 1573-2916.
\newblock \doi{10.1007/s10898-005-2454-3}.

\bibitem[Senn and Bretz(2007)]{senn_power_2007}
Stephen Senn and Frank Bretz.
\newblock Power and sample size when multiple endpoints are considered.
\newblock \emph{Pharmaceutical Statistics}, 6\penalty0 (3):\penalty0 161--170,
  July 2007.
\newblock ISSN 15391604, 15391612.
\newblock \doi{10.1002/pst.301}.

\bibitem[Roustant et~al.(2012)Roustant, Ginsbourger, and
  Deville]{roustant_dicekriging_2012}
Olivier Roustant, David Ginsbourger, and Yves Deville.
\newblock Dicekriging, diceoptim: Two r packages for the analysis of computer
  experiments by kriging-based metamodeling and optimization.
\newblock \emph{Journal of Statistical Software, Articles}, 51\penalty0
  (1):\penalty0 1--55, 2012.
\newblock ISSN 1548-7660.
\newblock \doi{10.18637/jss.v051.i01}.

\bibitem[Bossek et~al.(2020)Bossek, Doerr, and Kerschke]{bossek_initial_2020}
Jakob Bossek, Carola Doerr, and Pascal Kerschke.
\newblock Initial design strategies and their effects on sequential model-based
  optimization: An exploratory case study based on bbob.
\newblock In \emph{Proceedings of the 2020 Genetic and Evolutionary Computation
  Conference}, GECCO '20, pages 778--786, New York, NY, USA, June 2020.
  Association for Computing Machinery.
\newblock ISBN 978-1-4503-7128-5.
\newblock \doi{10.1145/3377930.3390155}.

\bibitem[Burnett and Jennison(2020)]{burnett_adaptive_2020}
Thomas Burnett and Christopher Jennison.
\newblock Adaptive enrichment trials: What are the benefits?
\newblock \emph{Statistics in Medicine}, pages 1--22, November 2020.
\newblock ISSN 0277-6715, 1097-0258.
\newblock \doi{10.1002/sim.8797}.

\bibitem[Stallard et~al.(2009)Stallard, Posch, Friede, Koenig, and
  Brannath]{stallard_optimal_2009}
Nigel Stallard, Martin Posch, Tim Friede, Franz Koenig, and Werner Brannath.
\newblock Optimal choice of the number of treatments to be included in a
  clinical trial.
\newblock \emph{Statistics in Medicine}, 28\penalty0 (9):\penalty0 1321--1338,
  2009.
\newblock ISSN 1097-0258.
\newblock \doi{10.1002/sim.3551}.

\end{thebibliography}

% \newpage
% \phantom{aaaa}

% %%%%%%%%%% Merge with supplemental materials %%%%%%%%%%
% \clearpage
% \begin{center}
% \textbf{\large Supplemental Materials: \newtitle}
% \end{center}
% \FloatBarrier
% %%%%%%%%%% Merge with supplemental materials %%%%%%%%%%
% %%%%%%%%%% Prefix a "S" to all equations, figures, tables and reset the counter %%%%%%%%%%
% \setcounter{equation}{0}
% \setcounter{figure}{0}
% \setcounter{table}{0}
% \setcounter{page}{1}
% \makeatletter
% \renewcommand{\theequation}{S\arabic{equation}}
% \renewcommand{\thefigure}{S\arabic{figure}}
% \renewcommand{\thetable}{S\arabic{table}}
% \renewcommand{\bibnumfmt}[1]{[S#1]}
% %%%%%%%%%% Prefix a "S" to all equations, figures, tables and reset the counter %%%%%%%%%%

% \input{generated/tables/table_best.tex}

% \begin{figure}[htb]
% \centering
% \includegraphics[width=\linewidth]{generated/figures/plot_opt_path.pdf}
% \caption{%
%   Optimization curves of each optimization run and the mean of all runs drawn as a black line.
%   }
% \label{fig:plot_opt_path} 
% \end{figure}

\end{document}